\documentclass[fleqn,twoside]{article}
\usepackage{espcrc2}
\usepackage{graphicx,amsmath,amscd,amssymb,times}
\newcommand{\Pol}{\mathcal{P}}
\newcommand{\w}{0.33\linewidth}

\title{Extended Rein--Sehgal model for tau lepton production}
\author{Konstantin S. Kuzmin%
\address[JINR]{Joint Institute for Nuclear Research,
               RU-141980, Dubna, Moscow region, Russia.}%
\address[ITEP]{Institute for Theoretical and Experimental Physics,
               RU-117218, Moscow, Russia.},
        Vladimir V. Lyubushkin%
\addressmark[JINR]%
\address[IrSU]{Physics Department, Irkutsk State University,
               RU-664003, Irkutsk, Russia.}
        and
        Vadim A. Naumov%
\addressmark[JINR]%
\address[INFN]{Dipartimento di Fisica, Universit\`a degli Studi di Firenze
               and INFN Sezione di Firenze, \\
               I-50019, Sesto Fiorentino (FI), Italy.}}
      
\begin{document}

\begin{abstract}
The polarization density matrix formalism is employed to
include the final lepton mass and spin into the popular model
by Rein and Sehgal for single pion neutrinoproduction.
We investigate the effect of the $\tau$ lepton mass on the 
differential cross sections. The lepton polarization evaluated
within the extended RS model is compared against that follows
from the single resonance production model based upon the
Rarita-Schwinger formalism with phenomenological transition
form factors.

\vspace{1pc}
\end{abstract}

\maketitle

\section{INTRODUCTION}
\label{sec:Introduction}

The Rein-Sehgal (RS) model~\cite{Rein:81,Rein:87} is undoubtedly
one of the most circumstantial and approved phenomenological tools
for description of single-pion production through baryon resonances
in neutrino and antineutrino interactions with nucleons.
It is incorporated into essentially all MC neutrino event generators
in the few-GeV region, developed for both accelerator and
astroparticle experiments~\cite{NeutrinoGenerators}.
However the RS model is not directly applicable to the $\nu_{\tau}$
and $\overline{\nu}_{\tau}$ induced reactions since it neglects
the final lepton mass. Due to the same reason, the model is not
suited for studying the lepton polarization phenomenon.

In Ref.~\cite{Kuzmin:03}, we proposed a generalization of the
RS model which takes into account the final lepton mass and spin;
it will be hereafter referred to as ``Extended Rein-Sehgal'' (ERS)
model.

In this paper we briefly summarize the key points of the ERS
model~\cite{Kuzmin:03} (Sect.~\ref{sec:PDM}) and discuss some
meaningful numerical results (Sect.~\ref{sec:Results}).

\section{SKETCH OF THE ERS MODEL}
\label{sec:PDM}

Our extension is based upon a covariant form of the charged leptonic
current $j_\lambda$ with definite lepton helicity $\lambda$, which
allows us to express the components $j_\lambda^\alpha$ of the current
in the resonance rest frame (RRF) through the kinematic variables
(and $\lambda$) measured in the laboratory frame. 
Since the leptonic current $j_\lambda$ still can be treated as the
intermediate $W$ boson polarization 4-vector, it may be decomposed
(in RRF) into three polarization 4-vectors $e_{L,R,S}$ corresponding
to left-handed, right-handed and scalar polarizations. 

However, the vector $e_S$ has to be modified with respect to that of
the original RS model and, consequently, its inner products with the
vector and axial charged hadronic currents $F^{V,A}$ have to be recalculated.
To do this, we used the explicit form for the currents $F^{V,A}$
\cite{Ravndal:73} of the Feynman-Kislinger-Ravndal (FKR) relativistic
quark model~\cite{Feynman:71} adopted in the RS approach.
As a result, the three structures, $S^V$, $B^A$ and $C^A$, involved
into the description of the FKR dynamics are also modified~\cite{Kuzmin:03}.

After that, the lepton polarization density matrix
$\boldsymbol{\rho}=\,\parallel\!\rho_{\lambda\lambda'}\!\parallel$
can be written as the superposition of the partial cross sections
$\sigma_i^{\lambda\lambda'}$ ($i=L,R,S$),
\begin{equation*}\label{rho}
\rho_{\lambda\lambda'}=\frac{\varSigma_{\lambda\lambda'}}
{\varSigma_{++}+\varSigma_{--}},
\quad
\varSigma_{\lambda\lambda'}=\sum\limits_{i=L,R,S}\!
c_i^{\lambda}c_i^{\lambda'}\sigma_i^{\lambda\lambda'},
\end{equation*}
and the differential cross section
is given by
\begin{equation*}\label{CS}
\frac{d^2\sigma}{dQ^2dW^2}=
\frac{G_F^2\cos^2\theta_CQ^2}{2\pi^2M\left|\mathbf{q}\right|^2}
\left(\varSigma_{++}+\varSigma_{--}\right).
\end{equation*}
The cross sections $\sigma_i^{\lambda\lambda'}$ are found to be
bilinear combinations of the CC amplitudes referring to one single
resonance in a definite state of isospin, charge and helicity.
These amplitudes remain \emph{the same} as in the original RS model.
The coefficients $c_i^{\lambda}$ are explicitly defined through
the components $j_\lambda^\alpha$ written in RRF.
The remaining kinematic variables and constants in the above
equation have their standard meaning.

\section{NUMERICAL RESULTS AND DISCUSSION}
\label{sec:Results}

In our calculations, we use the same set of 18th nucleon resonances
with masses below 2 GeV$/c^2$ as in Ref.~\cite{Rein:81} but with
all relevant parameters updated according to the most recent
data~\cite{Eidelman:04}.
The factors which were estimated in Ref.~\cite{Rein:81} numerically
are corrected by using the new data and a more accurate integration
algorithm. We do not modify the form of the transition form factors
$G^{V,A}\left(Q^2\right)$ adopted in the RS model but use the
today's standard values for the axial mass and coupling constant.

\begin{figure}[hbt]
\vskip  2mm
\includegraphics[width=\linewidth]{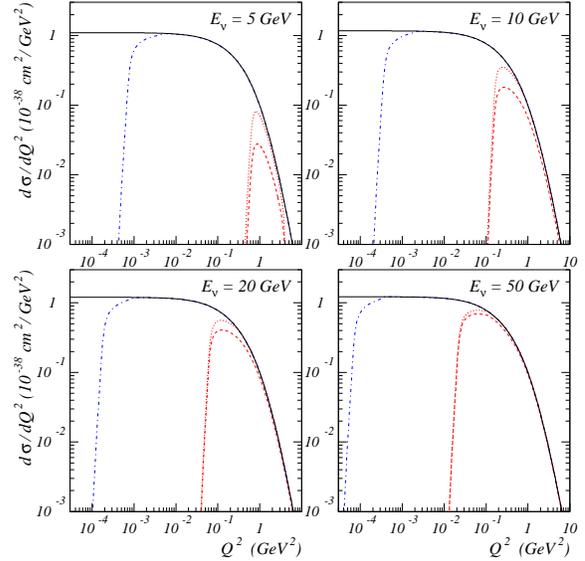}
\vskip -2mm
\caption{The finite lepton mass effect for
         the $\nu_{\tau}p\to\tau^-p\pi^+$ differential cross section
         at $E_\nu=5,10,20$ and 50 GeV.
         Solid and dotted lines are, respectively, for the standard
         RS model predictions with zero lepton mass and with the
         $\tau$ lepton mass included only into kinematics.
         Dashed lines are for the ERS model ($m_\tau$ is taken
         into account in both kinematics and dynamics).
         The $\nu_{\mu}p\to\mu^-p\pi^+$ cross section calculated
         in the ERS model is also plotted by dash-dotted lines.
         In all cases, the phase space has been cut by the
         condition $W<2$ GeV.
\label{fig:n_e05-50}}
\end{figure}

In Fig.~\ref{fig:n_e05-50} we show, as an example, the finite
lepton mass effect for the differential cross section $d\sigma/dQ^2$
of the reaction $\nu_{\tau}p\to\tau^-p\pi^+$. The calculations are
done for $W<2$ GeV. The major effect is, of course, due to the
$\tau$ lepton production threshold but the accounting for the mass
in the lepton current (``dynamic correction'') gives rise to a
significant additional decrease of the cross section:
the effect can be as large as 300\% at low neutrino energies and
remains important up to rather high energies.

For a comparison, in Fig.~\ref{fig:n_e05-50} we also show the
$\nu_{\mu}p\to\mu^-p\pi^+$ cross section calculated in the ERS
model. In this case, the dynamic mass correction is typically at
the few per cent level or less and the purely kinematic correction
for the muon mass is sufficient.
The described situation is qualitatively similar for the rest
$\nu$ and $\overline{\nu}$ induced reactions under consideration
so we do not show the corresponding figures.

\begin{figure*}[htb]
\vskip  2mm
\includegraphics[width=\w]{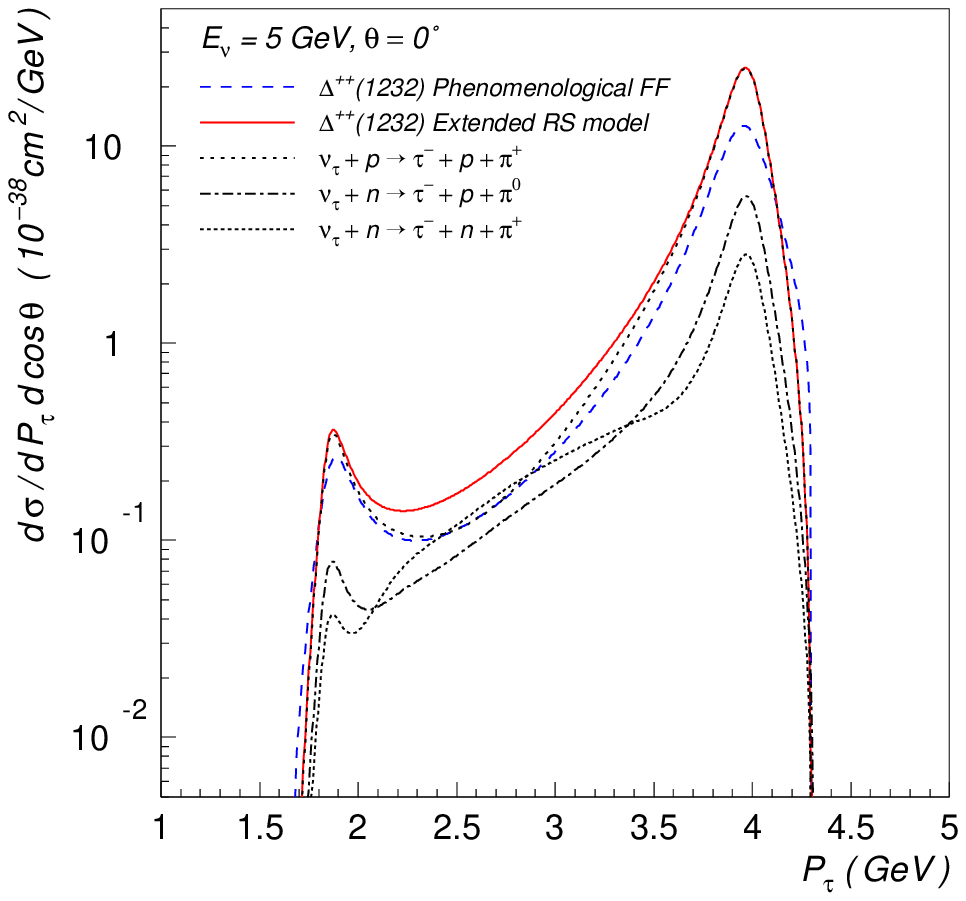}
\includegraphics[width=\w]{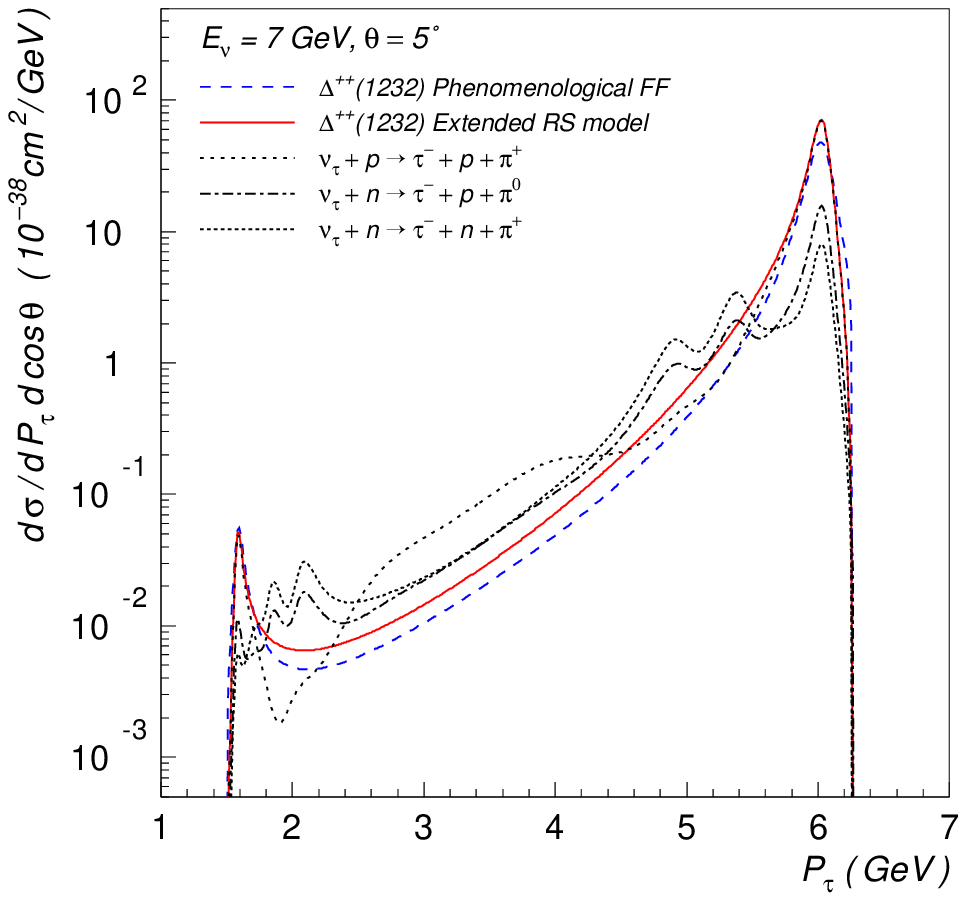}
\includegraphics[width=\w]{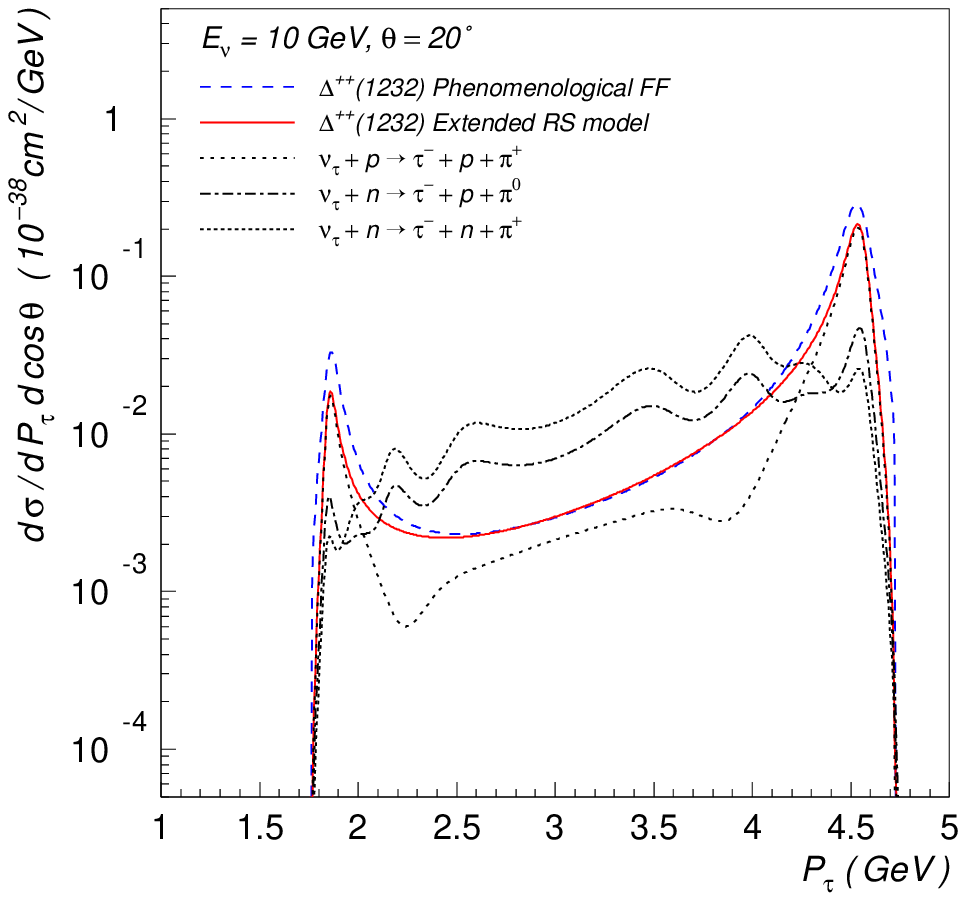}
\includegraphics[width=\w]{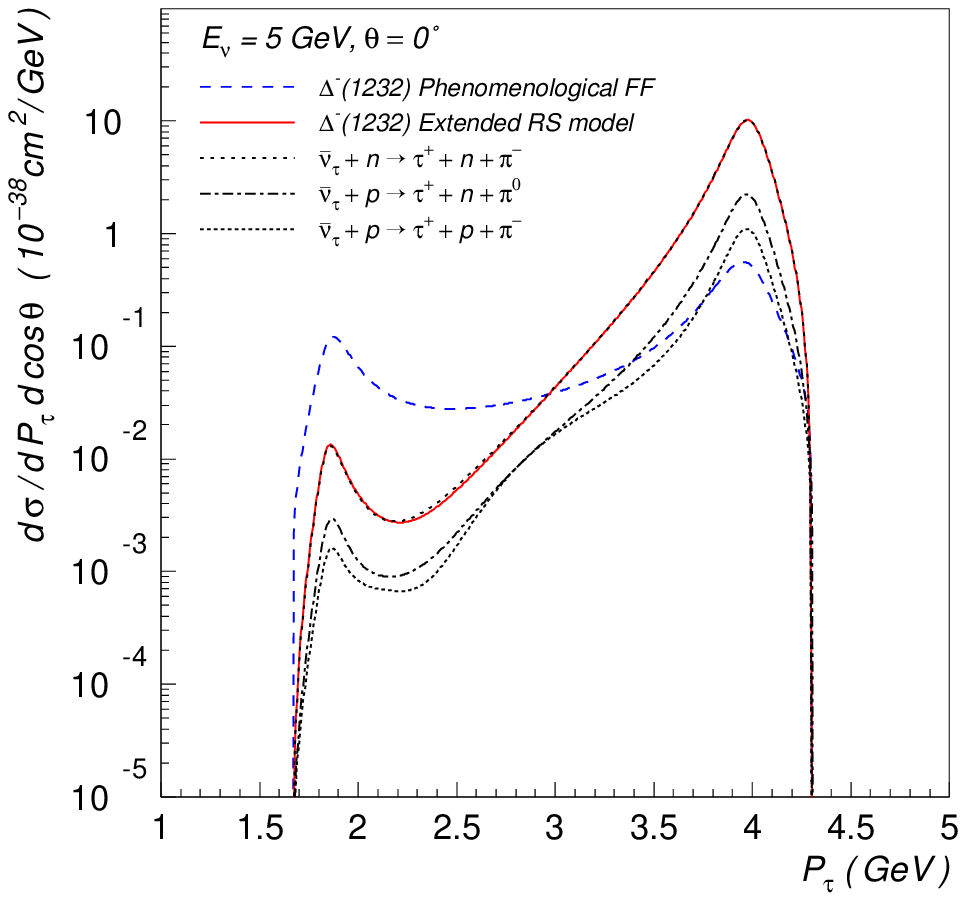}
\includegraphics[width=\w]{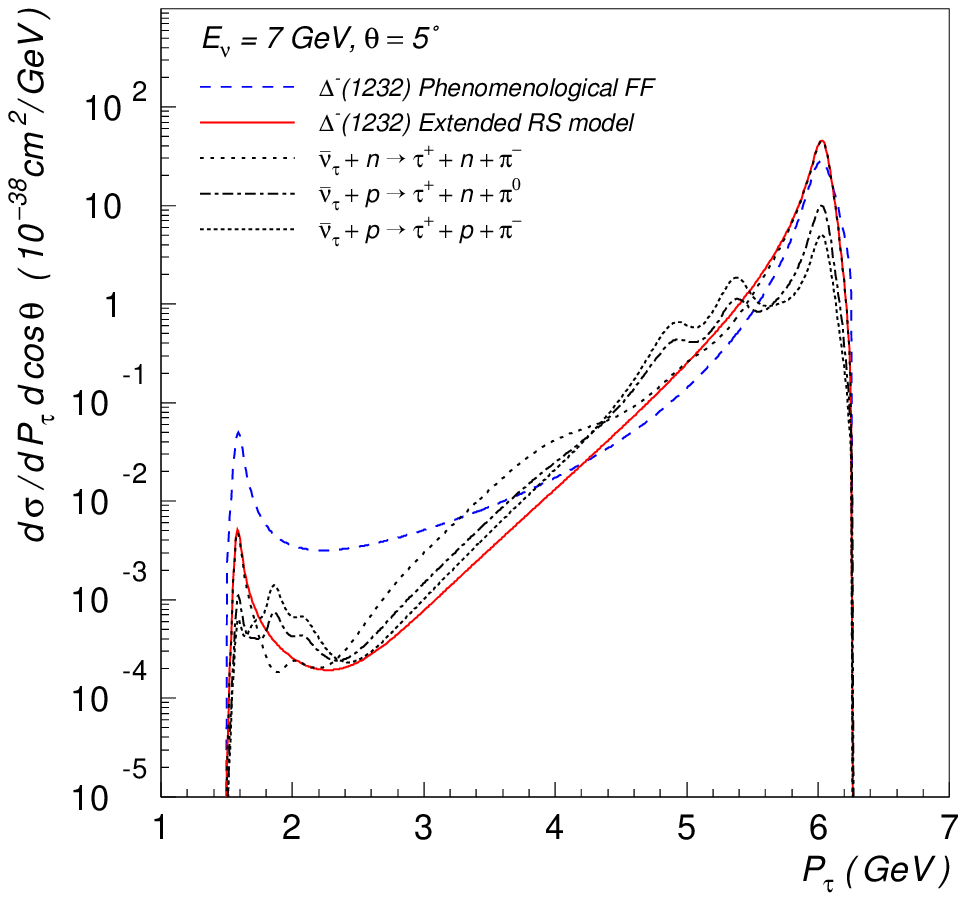}
\includegraphics[width=\w]{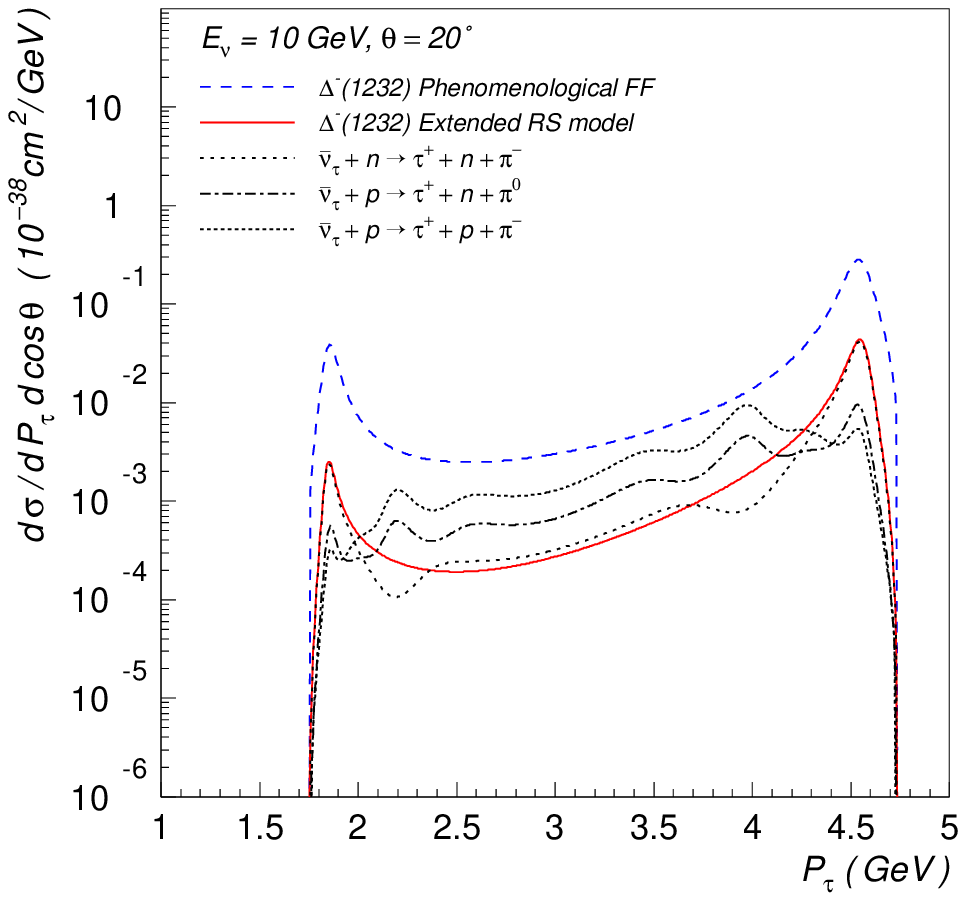}
\vskip -2mm
\protect\caption{Double differential cross sections
                 $d^2\sigma/dP_{\tau}d\theta$,
                 as functions of $\tau$ lepton momentum $P_\tau$ for
                 six reactions of single pion production by $\nu_\tau$
                 (top panels) and $\overline{\nu}_\tau$ (bottom panels).
                 Calculations are done in the ERS model.
                 The (anti)neutrino energies and scattering angles are
                 indicated in the legends. 
                 The cases for the single $\Delta(1232)$ resonance
                 production evaluated in the ERS model and in the
                 Rarita-Schwinger formalism with phenomenological
                 transition form factors are also plotted.
\label{f:Sigma}}
\end{figure*}

Figure~\ref{f:Sigma} shows the double differential cross sections
$d^2\sigma/dP_{\tau}d\theta$ as functions of $\tau$ lepton momentum
$P_\tau$ for six reactions of single pion production by $\tau$
neutrinos and antineutrinos. Calculations are done in the ERS model. 
The choice for the incident $\nu_\tau/\overline{\nu}_\tau$ energies
$E_\nu$ and the lepton scattering angles in the figure is quite
arbitrary but representative. The distinctive shape of the cross sections
with the two sharp Breit-Wigner cusps is a reflection of the two
kinematically allowed solutions (``branches'') for $P_\tau$ at fixed
values of $E_\nu$, $\theta$ and $W$.
So this feature is typical for $d^2\sigma/dP_{\tau}d\theta$ at
any allowed values of $E_\nu$ and $\theta$ and moreover it is much
the same for the individual resonance contributions into the cross section.
The amplitudes of the cusps and their ratio are however very responsive
to variations of both $E_\nu$ and $\theta$, e.g., near the reaction
threshold, the contribution of the left cusp into the total cross section
becomes important.

The fine structure of the cross sections is a result of the interference
of many contributions with very different magnitudes and shapes. Since the
resonance contributions are particular for each reaction, the fine
structure is also quite distinctive.

\begin{figure*}[htb]
\vskip  2mm
\includegraphics[width=\w]{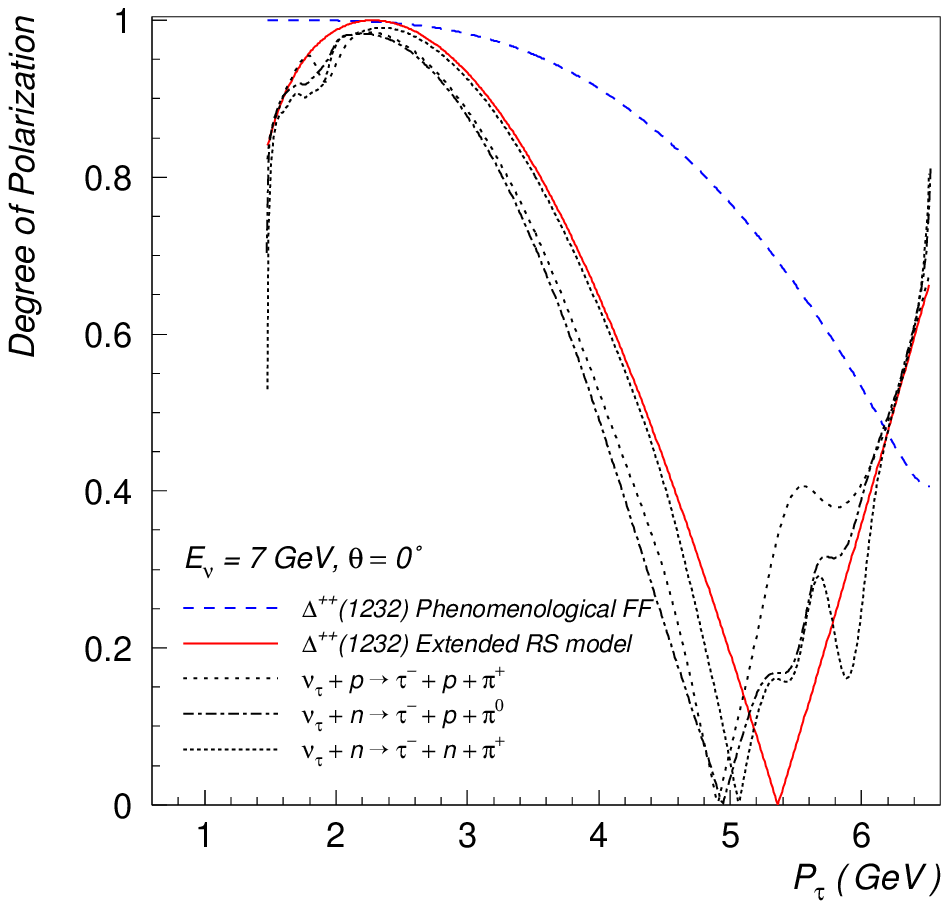}
\includegraphics[width=\w]{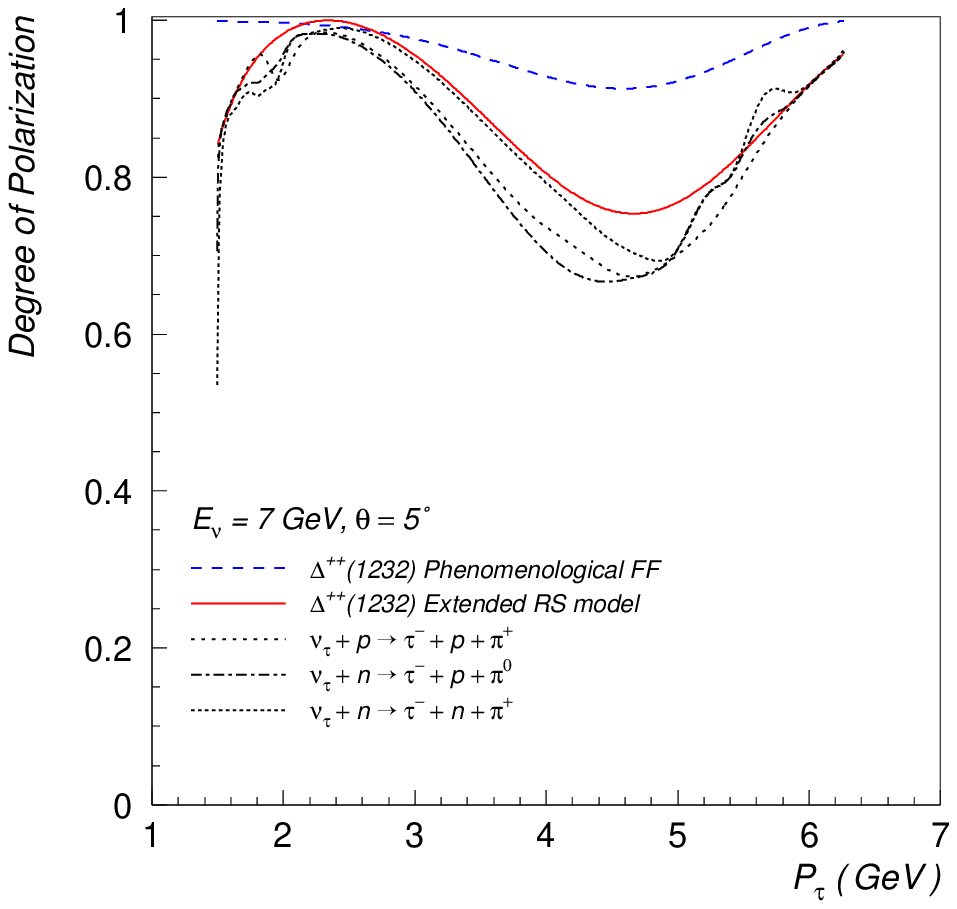}
\includegraphics[width=\w]{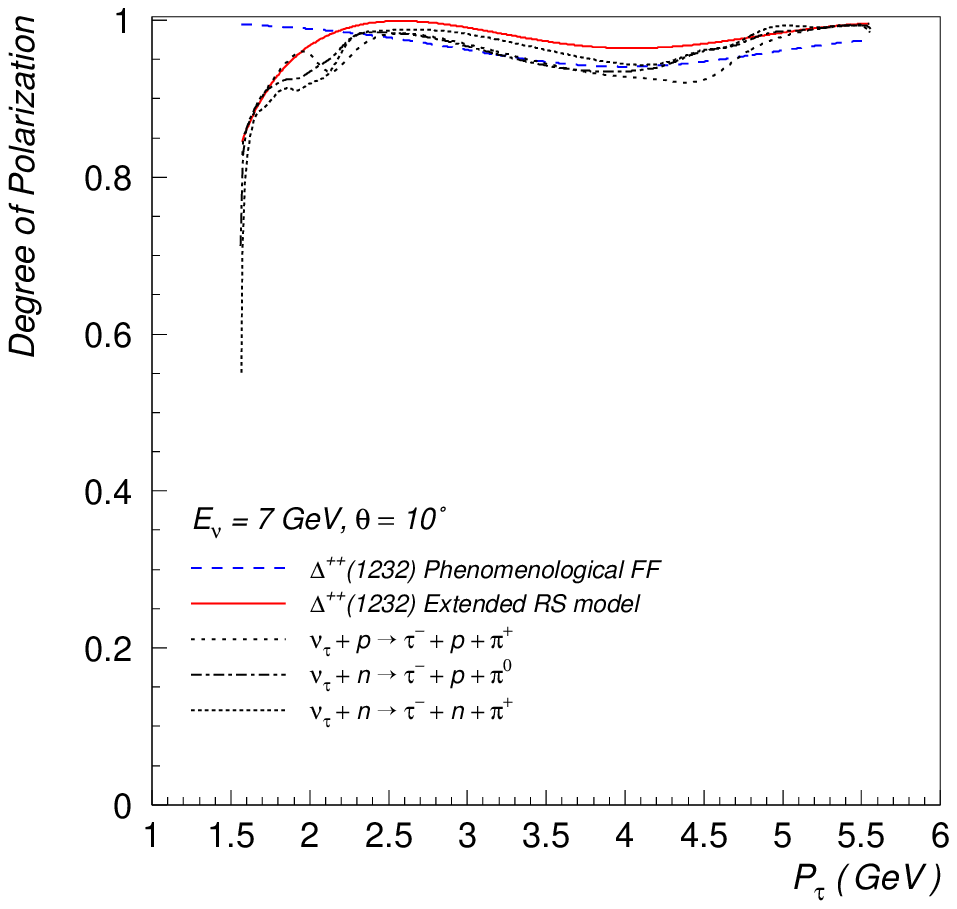}
\includegraphics[width=\w]{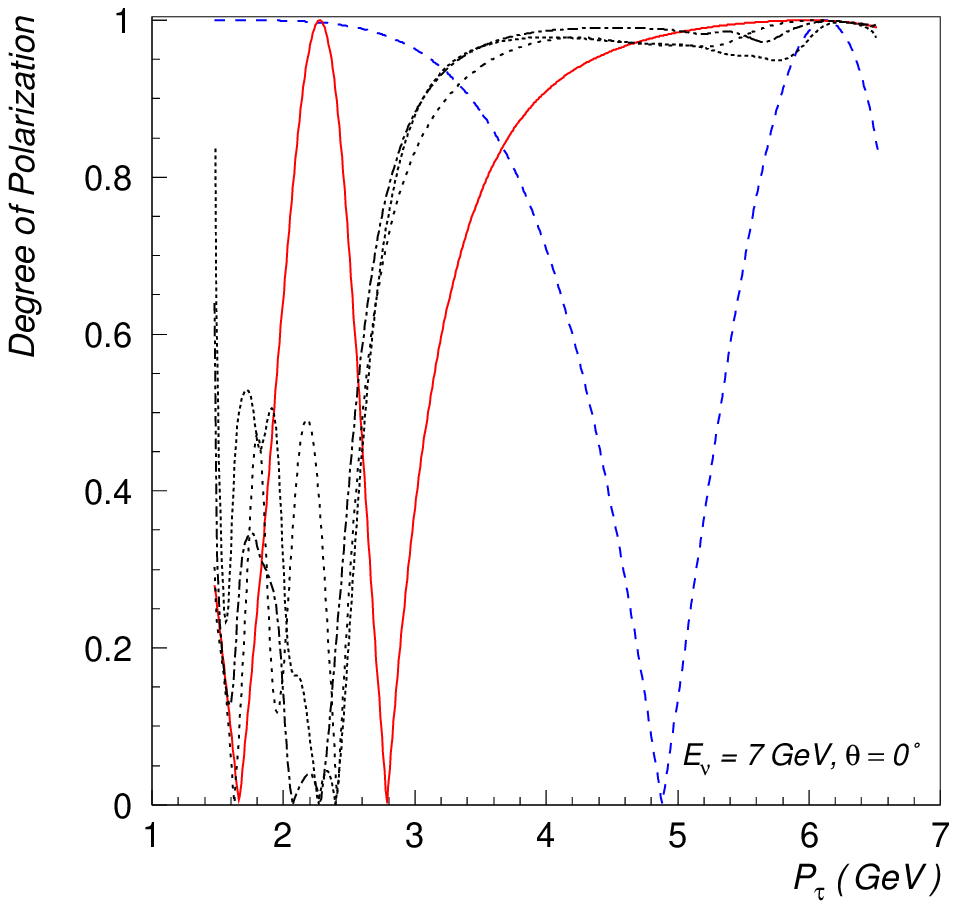}
\includegraphics[width=\w]{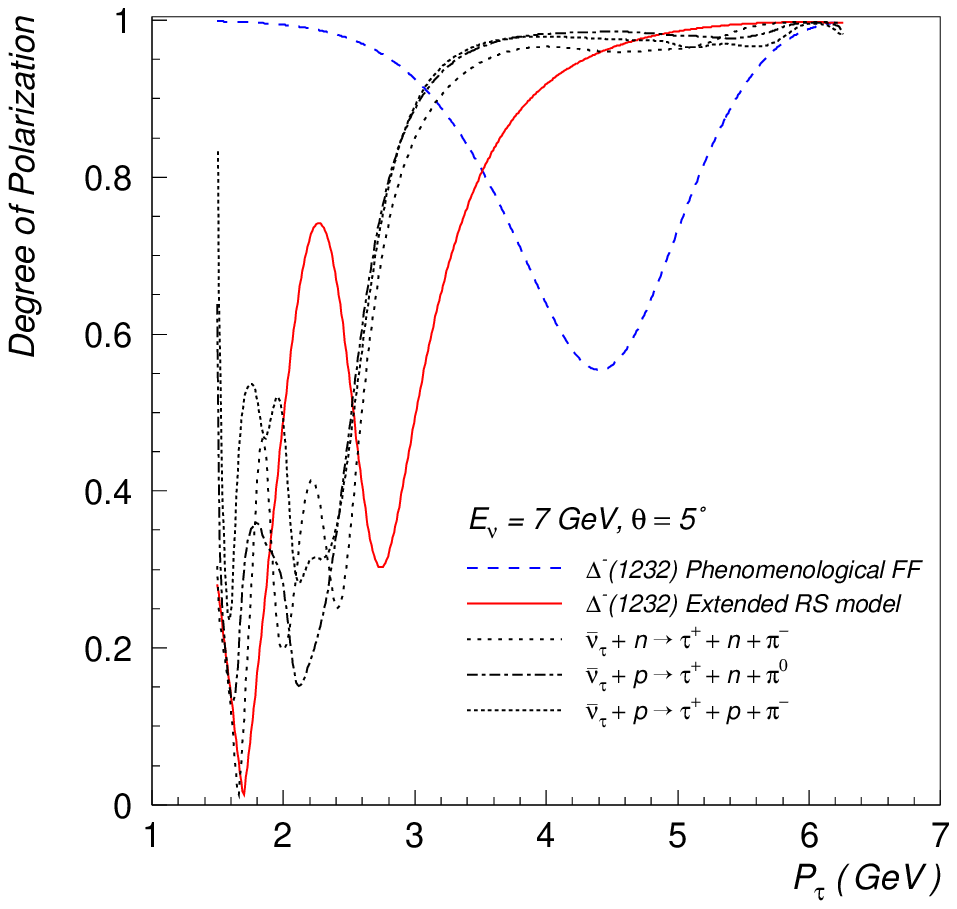}
\includegraphics[width=\w]{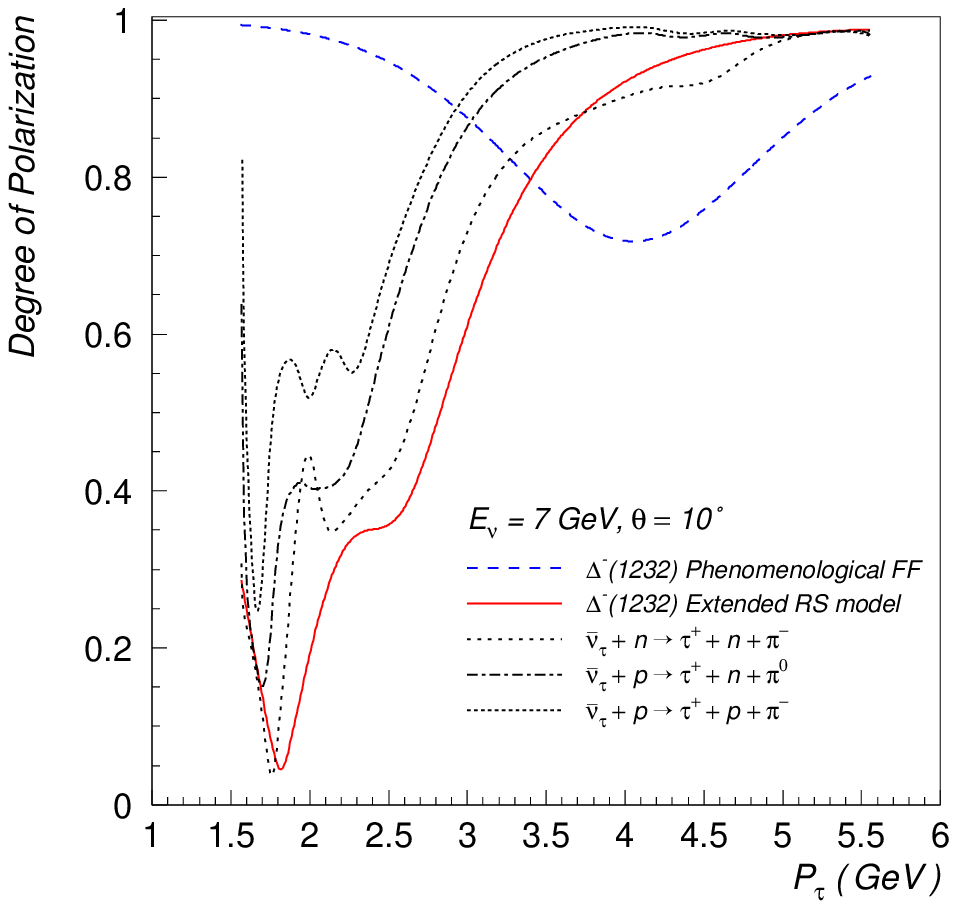}
\vskip -2mm
\protect\caption{Degree of polarization of $\tau$ leptons produced in
                 the $\nu_\tau$ and $\overline{\nu}_\tau$ induced reactions at
                 $E_\nu=7$ GeV for $\theta=0^\circ,5^\circ$ and $10^\circ$.
                 The cases for the single $\Delta(1232)$ resonance
                 production evaluated in the ERS model and in the
                 Rarita-Schwinger formalism with phenomenological
                 transition form factors are also plotted.
\label{f:P07}}
\end{figure*}

For comparison, the cases for the single $\Delta(1232)$ resonance
production evaluated in the ERS model and in the Rarita-Schwinger
formalism with phenomenological transition form factors as described
in Ref.~\cite{Kuzmin:03} are also shown. In the latter case we used
the same input parameters (borrowed from Ref.~\cite{Alvarez-Ruso:98})
as in the recent paper by Hagiwara et al.~\cite{Hagiwara:03}.

In Fig.~\ref{f:P07} we plot the degree of polarization
$\left|\boldsymbol{\Pol}\right|$ of $\tau$ leptons generated in the
same reaction and as in Fig.~\ref{f:Sigma} at three scattering angles:
$\theta=0^\circ,5^\circ$ and $10^\circ$. As an example, the incident
$\nu_\tau/\overline{\nu}_\tau$ energy $E_\nu$ is taken to be 7 GeV.
It can be seen from the figure that, for the three $\nu_\tau$ initiated
reactions, the behavior of $\left|\boldsymbol{\Pol}\right|$ is
qualitatively the same for any lepton momentum. 
For the three reactions initiated by $\overline{\nu}_\tau$,
the shapes of $\left|\boldsymbol{\Pol}\right|$ are qualitatively
comparable in the region of the main kinematic branch
(the neighborhood of the right cusps in Fig.~\ref{f:Sigma}) but
become drastically different in the neighborhoods of the left cusps. 
The distinction of the individual profiles of $\left|\boldsymbol{\Pol}\right|$
for different reactions is again a result of the interference of
different groups of the resonances with different amplitudes.

For comparison, the case with the single $\Delta(1232)$ resonance
production is also shown in Fig.~\ref{f:P07}. In this case,
the degree of polarization was evaluated in the ERS model and by
using the Rarita-Schwinger formalism with the phenomenological transition
form factors suggested in Ref.~\cite{Alvarez-Ruso:98}.
As for the differential cross sections, we used the same set of
inputs as in Ref.~\cite{Hagiwara:03}. Our results are rather close to
those of Ref.~\cite{Hagiwara:03} but the agreement is not complete;
in particular, Hagiwara et al.\ neglected the second kinematic solution.
It is however more important that the predictions of the ERS model and
the Rarita-Schwinger approach for the single $\Delta(1232)$ production
are distinctly different for $\nu_\tau$ initiated reactions
(especially for small scattering angles and large lepton momenta)
and completely different for $\overline{\nu}_\tau$ initiated reactions
(essentially everywhere).

Qualitatively similar conclusions can also be drawn for the longitudinal
and perpendicular components of the polarization vector
not illustrated in this paper.

\end{document}